\documentclass[amsmath,amssymb,,mathbbm,superscriptaddress,floatix]{revtex4}
\usepackage{amsmath,amsfonts,amssymb,amsthm,graphics,graphicx,epsfig,bbm}
\usepackage[colorlinks=true,citecolor=blue,linkcolor=blue,urlcolor=blue]{hyperref}
\usepackage[usenames]{color}
\usepackage{subfigure}
\usepackage{epsfig}
\usepackage{dcolumn}
\usepackage{bm}
\usepackage{color}
\usepackage{verbatim}
\usepackage{epstopdf}
\usepackage{amstext}
\usepackage{latexsym}
\usepackage{hyperref}
\usepackage{psfrag}
\usepackage{xcolor}
\usepackage{times}
\usepackage{bbold}

\begin{document}

\title{The Forbidden Quantum Adder}

\author{U. Alvarez-Rodriguez} \email{unaialvarezr@gmail.com}

\affiliation{Department of Physical Chemistry, University of the Basque Country UPV/EHU, Apartado 644, E-48080 Bilbao, Spain}
\author{M. Sanz}
\affiliation{Department of Physical Chemistry, University of the Basque Country UPV/EHU, Apartado 644, E-48080 Bilbao, Spain}
\author{L. Lamata}
\affiliation{Department of Physical Chemistry, University of the Basque Country UPV/EHU, Apartado 644, E-48080 Bilbao, Spain}
\author{E. Solano}
\affiliation{Department of Physical Chemistry, University of the Basque Country UPV/EHU, Apartado 644, E-48080 Bilbao, Spain}
\affiliation{IKERBASQUE, Basque Foundation for Science, Maria Diaz de Haro 3, 48013 Bilbao, Spain}

\begin{abstract}
Quantum information provides fundamentally different computational resources than classical information. We prove that there is no unitary protocol able to add unknown quantum states belonging to different Hilbert spaces. This is an inherent restriction of quantum physics that is related to the impossibility of copying an arbitrary quantum state, i.e., the no-cloning theorem. Moreover, we demonstrate that a quantum adder, in absence of an ancillary system, is also forbidden for a known orthonormal basis.  This allows us to propose an approximate quantum adder that could be implemented in the lab. Finally, we discuss the distinct character of the forbidden quantum adder for quantum states and the allowed quantum adder for density matrices.
\end{abstract}

\date{\today}

\maketitle

Addition plays a central role in mathematics and physics, while adders are ubiquitous devices in the fields of computation~\cite{ComputationBook} and electronics~\cite{ElectronicsBook}. In this sense, usual sum operations can be realized by classical Turing machines~\cite{Turing} and also, with a suitable algorithm, by quantum Turing machines~\cite{Barenco,Deutsch}. Furthermore, the sum of known state vectors in the same Hilbert space, i.e. quantum superposition, is at the core of quantum physics. In fact, entanglement and the promised exponential speed-up of quantum computing are based on such linear combinations. Here, we consider the existence of a quantum adder, defined as a unitary operation mapping two unknown quantum states encoded in different quantum systems onto their sum codified in a single system. The surprising answer is that this quantum adder is forbidden and it has the quantum cloner as a special case~\cite{WoottersZurek}. This no-go result, as other prohibited operations~\cite{WoottersZurek,noprog,nopo,noref}, is of fundamental nature and its implications should be further studied. Furthermore, we consider a high-fidelity approximate quantum adder involving ancillary systems. Recently, we have known about a parallel work analyzing a similar problem, in which an optimal approximate quantum adder was found~\cite{sup}.

\begin{figure}[h!!]
\begin{center}
\includegraphics[width=9cm, trim=3cm 0cm 0cm 3cm]{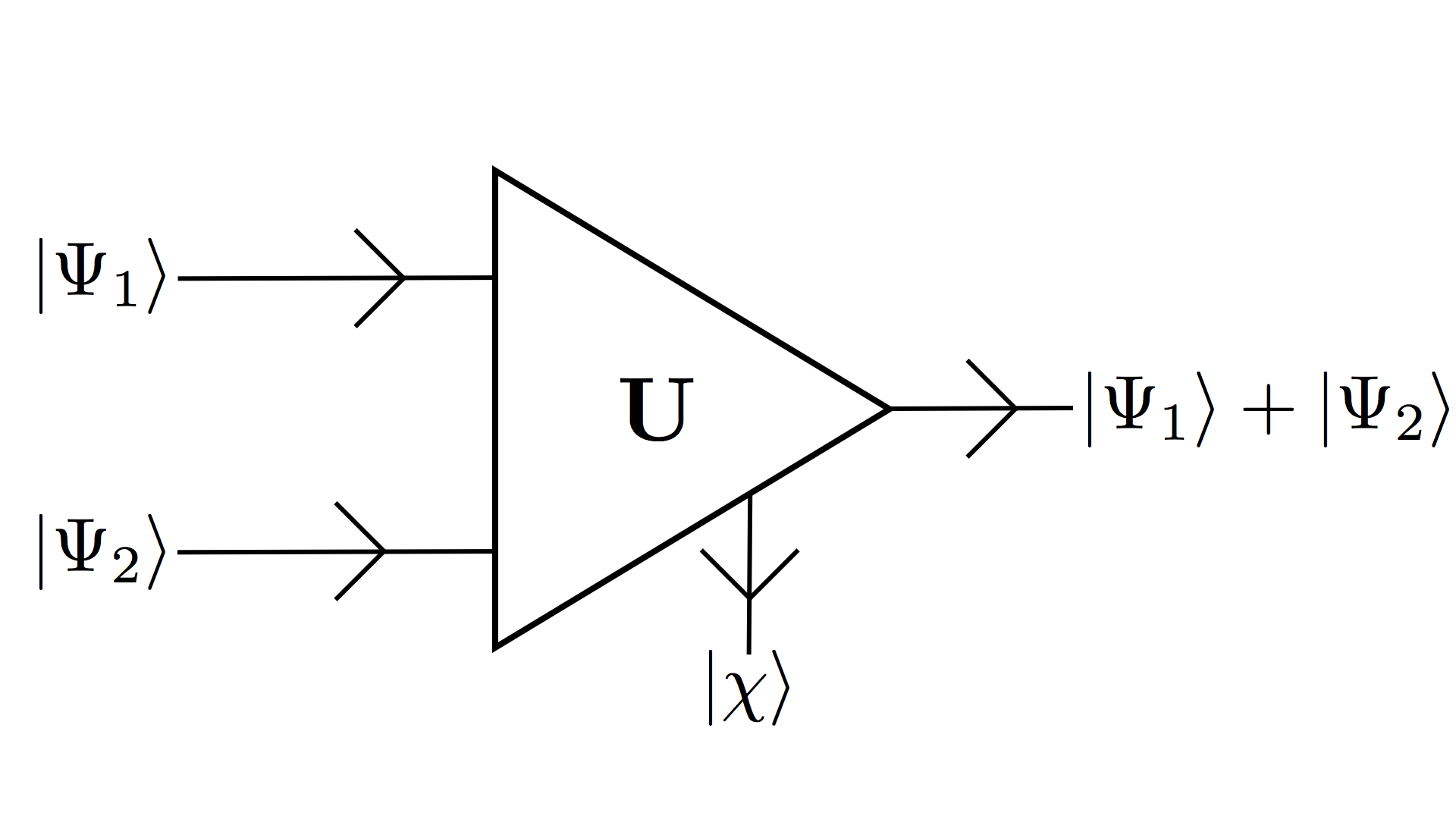}
\caption{ Scheme of the conjectured quantum adder. The inputs are two unknown quantum states, $|\Psi_1\rangle$ and $|\Psi_2\rangle$, while the outputs are proportional to the sum, $|\Psi_1\rangle + |\Psi_2\rangle$ with an ancillary state $|\chi\rangle$.}
\label{Fig1}
\end{center}
\end{figure}

Let $|\Psi_1 \rangle, |\Psi_2 \rangle \in \mathbb{C}^d$ be two quantum states of a finite-dimensional Hilbert space. The conjectured quantum adder, sketched in Fig.~\ref{Fig1}, would be a mathematical operation defined as the unitary $U: \mathbb{C}^d\otimes \mathbb{C}^d \rightarrow \mathbb{C}^d\otimes \mathbb{C}^d$, for every pair of unknown $| \Psi_1\rangle \otimes |\Psi_2\rangle$ and ancillary vector $|\chi\rangle \in\mathbb{C}^d$,
\begin{equation}
\label{sum}
U |\Psi_1 \rangle |\Psi_2 \rangle \propto ( |\Psi_1 \rangle + |\Psi_2 \rangle) |\chi \rangle ,
\end{equation}
where the ancillary state $| \chi \rangle$ may depend on the input states. There are several ways of proving the unphysicality of Eq.~(\ref{sum}). The simplest one is to note that the unobservable global phase on its l.h.s. \!\!could be distributed in infinite forms on its r.h.s., $U e^{i \phi} |\Psi_1 \rangle |\Psi_2 \rangle = U e^{i \phi_1} |\Psi_1 \rangle e^{i \phi_2} | \Psi_2 \rangle \propto ( e^{i \phi_1} |\Psi_1 \rangle + e^{i \phi_2} |\Psi_2 \rangle) |\chi \rangle$, with $\phi = \phi_1 + \phi_2$, yielding an observable relative phase. When the ancillary state $|\chi\rangle$ does not depend on the input quantum states, the (forbidden) quantum cloner becomes a particular case of this restricted quantum adder. This follows from applying $U$ to two equal state vectors $U |\Psi\rangle |\Psi \rangle = |\Psi\rangle |\chi\rangle$, since the inverse generates a quantum cloning operation. Therefore, although the general case of the quantum adder is not equivalent to a quantum cloner, it is still forbidden.

We consider now a different question, whether a quantum adder may exist for a given orthonormal basis. In this case, as we will see, the global phase does not produce any ambiguity in the equations. Let us consider the action of the unitary operator $U$ onto a set of orthonormal vectors: $U |0\rangle \,|0\rangle = | 0\rangle \, |B_0 \rangle$, $U |0\rangle \,|i\rangle = \frac{1}{\sqrt{2}}(|0\rangle + |i\rangle) \, |B_i \rangle$ and $U |i\rangle \,|0\rangle = \frac{1}{\sqrt{2}}(|0\rangle + |i\rangle) \, |\tilde{B}_i \rangle$, with $i = 1, \ldots , d-1$. Hence, as $U$ is a unitary matrix, it imposes some orthogonality conditions on the final vectors, $\langle B_0 | B_i\rangle  =\langle B_0|\tilde{B_j}\rangle=\langle B_i|\tilde{B_j}\rangle =0$ and $\langle B_i | B_j\rangle = \langle \tilde{B_i}|\tilde{B_j}\rangle= \delta_{i, j}$, with $i,j = 1, \ldots , d-1$. The second subspace has dimension $d$, but these constraints require the existence of at least $2d-1$ orthonormal vectors, which is impossible.

\begin{figure}[h!!]
\begin{center}
\includegraphics[width=\textwidth, trim=0cm 0cm 3cm 0cm]{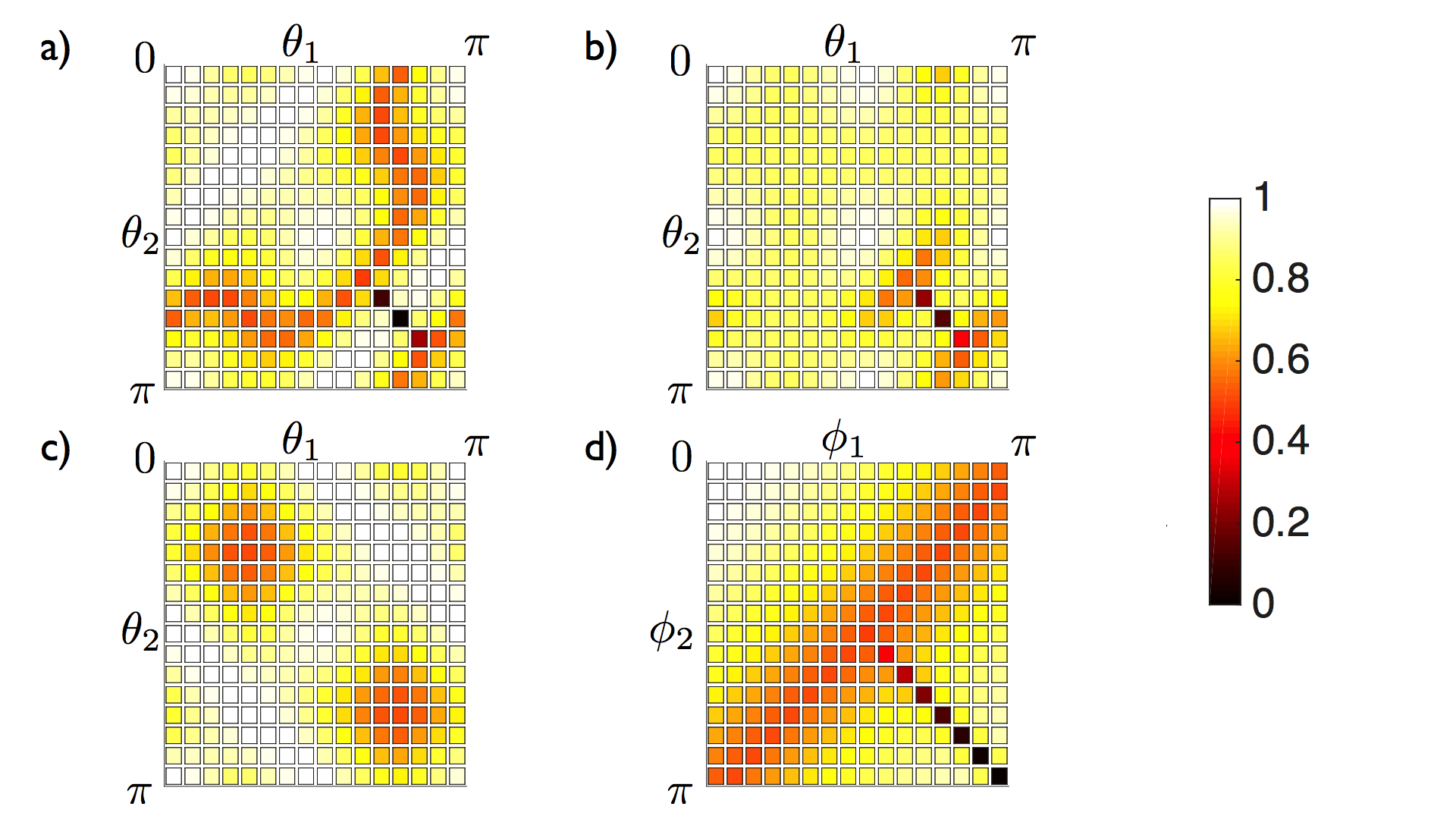}
\caption{Fidelity of the proposed approximate quantum adder as a function of the parameters of the input states $|\Psi_j\rangle=\cos \theta_j |0\rangle + \sin \theta_j e^{i \phi_j} |1\rangle$, where $j = 1, 2$. Here, a)~$\phi_1=\phi_2=0$, b)~$\phi_1=\phi_2=\pi/4$, c)~$\phi_1=\phi_2=\pi/2$, and d)~$\theta_1=\theta_2=\pi/4$. Note that the diagonal line of each plot corresponds to the approximate quantum cloner that is related to our restricted quantum adder. In this case, the fidelities are the lowest.}
\label{Fig2}
\end{center}
\end{figure}

We propose now the use of an ancillary system $|A\rangle$, which will assure the physicality and experimental feasibility of an approximate quantum adder for arbitrary unknown quantum states, see Fig.~\ref{Fig2}. This particular adder, $U_a$, computes the exact sum of the basis elements in qubit systems. Moreover, $U_a$ is extended by linearity to the whole Hilbert space, and implements an approximate sum when the input states are superpositions of the basis elements. The adder is given by the following expression in which $|B_i\rangle$ are orthonormal and $|+\rangle=\frac{1}{\sqrt{2}} (|0\rangle + |1\rangle)$. 
\begin{eqnarray}
&& U_a |0\rangle|0\rangle|A\rangle = |0\rangle|B_1\rangle, \qquad  \, U_a |0\rangle|1\rangle|A\rangle = |+\rangle|B_2\rangle, \\  && \nonumber U_a|1\rangle|0\rangle|A\rangle = |+\rangle|B_3\rangle, \qquad  U_a|1\rangle|1\rangle|A\rangle=|1\rangle|B_1\rangle.
\end{eqnarray}

Although this approximate quantum adder shows high fidelities, it is not optimal as the one recently found in Ref.~\cite{sup}.

Beyond the sum of quantum states in Eq.~\eqref{sum}, we may also consider the statistical addition of density matrices. Here, the input states are the tensor product of any pair of density matrices $\sigma = \rho_1 \otimes \rho_2 \in \mathcal{B}(\mathbb{C}^{2d} \otimes  \mathbb{C}^{2d})$, while the output state is the statistical sum $\rho = \frac{1}{2}(\rho_1 + \rho_2)$. The Kraus operators of the quantum channel realizing this adder are given by $E_i=\frac{1}{\sqrt{2}} (\mathbb{1}_{d} \otimes \langle i |)$ and $F_j=\frac{1}{\sqrt{2}} (\langle j | \otimes \mathbb{1}_{d})$, with $1 \leq i,j \leq d$. These operators straightforwardly perform the sum, i.e., ${\mathcal E}(\sigma) = \sum^{d}_{k=1} E_{k} \sigma E^{\dag}_k + \sum^{d}_{k=1} F_{k} \sigma F^{\dag}_k  = \frac{1}{2}(\rho_1 + \rho_2)$. Moreover, properly modified Kraus operators allow us to extend the previous result to any convex combination of input states. Therefore, the considered addition of density operators is always possible.

Let us compare the adders for state vectors and density operators. By writing the input states in Eq.~\eqref{sum} as density matrices, $\rho = |\Psi_1\rangle\langle \Psi_1|\otimes|\Psi_2\rangle\langle \Psi_2|$, both adders yield
\begin{eqnarray}
U\rho U^\dag &\propto& |\Psi_1\rangle\langle \Psi_1| + |\Psi_1\rangle\langle \Psi_2|+|\Psi_2\rangle\langle \Psi_1|+|\Psi_2\rangle\langle \Psi_2|, \nonumber \\ {\mathcal E}(\rho) &=& \frac{1}{2}(|\Psi_1\rangle\langle \Psi_1| + |\Psi_2\rangle\langle \Psi_2|).
\label{comp}
\end{eqnarray}
By comparing the adders in Eq.~\eqref{comp}, we can infer that the one in Eq.~\eqref{sum} would require the knowledge of the sum coherences, which were supposed unknown. This prohibition is of fundamental character in quantum physics, comparable and deeply related to the no-cloning theorem.

\vspace{1cm}

{\setlength{\parindent}{0pt}

{\bf Acknowledgments}
The authors acknowledge support from Spanish MINECO FIS2012-36673-C03-02; Ram\'on y Cajal Grant RYC-2012-11391; UPV/EHU UFI 11/55 and EHUA14/04, Basque Government IT472-10 and BFI-2012-322; CCQED, PROMISCE, SCALEQIT EU projects.\\

{\bf Author contributions}
U. A.-R. made the calculations while U. A.-R., M. S., L. L. and E. S. developed the protocol and wrote the manuscript. \\

{\bf Additional information}
The authors declare no competing financial interests. 

\end{document}